\documentclass{iaus}
\usepackage{graphicx}
\def\micron{$\mu$m}

\def\msun{\ifmmode M_{\odot} \else M$_{\odot}$\fi}
\def\msunyr{\ifmmode M_{\odot} {\rm yr}^{-1} \else M$_{\odot}$ yr$^{-1}$\fi}
\def\zsun{\ifmmode Z_{\odot} \else Z$_{\odot}$\fi}
\def\lsun{\ifmmode L_{\odot} \else L$_{\odot}$\fi}

\newcommand{\ha}{\ifmmode {\rm H}\alpha \else H$\alpha$\fi}
\newcommand{\hb}{\ifmmode {\rm H}\beta \else H$\beta$\fi}
\newcommand{\lya}{\ifmmode {\rm Ly}\alpha \else Ly$\alpha$\fi}

\newcommand{\ebv}{\ifmmode E_{\rm B-V} \else $E_{\rm B-V}$ \fi}
\def\zphot{\ifmmode z_{\rm phot} \else $z_{\rm phot}$\fi}
\def\bacs{B$_{\rm 435}$}
\def\vacs{V$_{\rm 606}$}
\def\iacs{i$_{\rm 776}$}
\def\zacs{z$_{\rm 850LP}$}

\def\lsim{\lower.5ex\hbox{\ltsima}~}
\def\gtsima{$\buildrel>\over\sim$}
\def\ga{\lower.5ex\hbox{\gtsima}~}

\title[~~Importance of nebular emission on the SED of high-z galaxies] 
{The importance of nebular emission for SED modeling of distant star-forming galaxies}

\author[D. Schaerer \& S. de Barros]   
{Daniel Schaerer$^{1,2}$
\and
Stephane de Barros$^1$}

\affiliation{$^1$Observatoire de Gen\`eve, Universit\'e de Gen\`eve, 51 Ch. des Maillettes, 1290 Versoix, Switzerland
 \\ email: {\tt daniel.schaerer@unige.ch} \\[\affilskip]
$^2$CNRS, IRAP, 14 Avenue E. Belin, 31400 Toulouse, France}

\pubyear{2011} 
\volume{284}  
\pagerange{1--12}
\jname{The Spectral Energy Distribution of Galaxies}
\editors{R.J. Tuffs \&  C.C.Popescu, eds.}
\begin{document}

\maketitle

\begin{abstract}
We highlight and discuss the importance of accounting for nebular emission in the SEDs of 
high redshift galaxies, as lines and continuum emission can contribute significantly or subtly 
to broad-band photometry. 
Physical parameters such as the galaxy age, mass, star-formation rate, dust attenuation and others
inferred from SED fits can be affected to different extent by the treatment of nebular emission.

We analyse a large sample of Lyman break galaxies from $z \sim$ 3--6, and show some
main results illustrating e.g.\ the importance of nebular emission for determinations of the mass--SFR relation,
attenuation and age. We suggest that a fairly large scatter in such relations could be intrinsic.
We find that the majority of objects ($\sim$ 60--70\%) is better fit with SEDs accounting 
for nebular emission; the remaining galaxies are found to show relatively weak or no 
emission lines. Our modeling, and supporting empirical evidence, suggests the existence of 
two categories of galaxies, ``starbursts'' and ``post-starbursts'' (lower SFR and older galaxies)
among the LBG population, and relatively short star-formation timescales.

\end{abstract}

\firstsection 
\section{Introduction}

Until recently the importance of nebular emission on the SED of high redshift galaxies and hence on 
on our understanding of the physical parameters of these objects was overlooked. For example, when
Spitzer started to detect the rest-frame optical part of the spectrum of high-redshift galaxies  redshifted
into broad-band filters of its IRAC camera at 3.6 and 4.5 \micron, it was soon noted that some of these
galaxies showed fairly large Balmer breaks \cite[(e.g.\ Egami et al.\ 2005, Eyles et al.\ 2005)]{Egami05,Eyles05}.
Several authors interpreted the large break as being due
to the presence of ``old", evolved stellar populations with ages of several hundred Myr 
\cite[(Eyles et al.\ 2005, Labbe et al.\ 2010)]{Eyles05,Labbe10}. 
If true this would imply the presence of very early galaxy populations, and very large formation redshifts 
($z_f \sim 10-30$). 

However, since emission lines are ubiquituous in star-forming galaxies and since atomic physics causes 
most of the strong observed lines to be in the (rest-frame) optical part of the spectrum, longward of the Balmer
break, nebular emission affects  broad-band photometry differentially, and can cause an apparent
Balmer break. The importance of this effect for high-z galaxies was pointed out by \cite[Schaerer \& de Barros (2009)]{SdB09},
who showed that significantly younger ages could be obtained when accounting for nebular lines.

It has now become clear \cite[(Schaerer \& de Barros 2010, Ono et al.\ 2010, Lidman et al.\ 2011)]{SdB09,Ono,Lidman11} 
that nebular emission (both lines and continuum emission) must be taken
into account for the interpretation of photometric measurements of the SEDs of star-forming galaxies such
as Lyman-alpha emitters (LAE)  and Lyman break galaxies (LBGs) -- the dominant galaxy populations at high-z.

Furthermore, as testified by the presence of \lya\ emission, a large (and possibly growing) fraction of 
the currently know population of star-forming galaxies at high redshift shows emission lines 
\cite[(at least \lya !; cf.Ouchi et al.\ 2008, Stark et al.\ 2010 and others)]{ouchi08,stark10}.

In parallel a lot of diverse evidence for galaxies with strong emission lines and/or strong contributions 
of nebular emission to broad-band filters has been found at different redshifts, e.g.\ by
\cite[Shim et al.\  (2011), McLinden et al.\ (2011), Atek et al.\ (2011), Trump et al.\ (2011), van der Wel 
et al.\ (2011)]{shimetal11,McLinden11,Atek11,Trump11,Vanderwel11}.

Here we briefly summarise our method to account for nebular emission in a SED plus photometric-redshift  
fitting code and present some results obtained from a systematic study of a large sample of $z \sim$ 3--6 
LBGs.

\section{SED modeling}
Several evolutionary synthesis models, such as P\'EGASE, GALEV and others 
include nebular emission  \cite[(cf. Fioc et al.\ 1999, Charlot \& Longhetti 2001, Anders \& Fritze 2003, 
Zackrisson et al.\ 2008)]{fioc99,Charlot01,Anders03,Zackrisson08}.
However, they were not used to analyse observations of high redshift galaxies.

Our SED fitting tool, described in \cite[Schaerer \& de Barros (2009, 2010)]{SdB09,SdB10},
is based on a version of the {\em Hyperz} photometric redshift code of \cite[Bolzonella et al.\ (2000)]{hyperz}, 
modified to taking into account nebular emission.
In \cite[de Barros et al.\ (2011, hereafter dB11)]{dB11} we consider a large set of spectral templates based on
the GALAXEV synthesis models of \cite[Bruzual \& Charlot (2003)]{BC03}, covering
different metallicities and a wide range of star formation
histories. A Salpeter IMF is adopted.
Nebular emission from continuum processes and numerous emission lines is added to the
spectra predicted from the GALAXEV models as described in
\cite[Schaerer\& de Barros (2009)]{SdB09}, proportionally to the Lyman continuum
photon production. 
The intergalactic medium (IGM) is treated with the prescription of \cite[Madau (1995)]{Madau95}.

The free parameters of our SED fits are:
redshift $z$,  metallicity $Z$ (of stars and gas),  
star formation (SF) history described by the timescale $\tau$ (i.e.\ SFR
$\propto \exp(-t/\tau)$),
the age $t_\star$ defined since the onset of star-formation, and
attenuation $A_V$ described by the Calzetti law.
In some models we also vary the strength of the \lya\ line  from zero to its maximum
predicted by case B \cite[(see Schaerer et al.\ 2011)]{Schaerer11}.
In \cite[dB11]{dB11} we consider 3 sets of SF histories, exponentially declining SF with variable timescales,
 SFR=const, and rising star-formation following \cite[Finlator et al.\ (2011)]{Finlator11}.
We use Monte-Carlo simulations to determine probability distribution functions for each 
parameter/quantity.

\section{Impact of nebular emission on physical parameters of star-forming galaxies from $z \sim 3$ to 8}

In \cite[Schaerer \& de Barros (2009)]{SdB09} we have demonstrated the effect nebular
emission may have on age determinations of high-z galaxies.
In \cite[Schaerer \& de Barros (2010)]{SdB10} we have applied our SED fitting tool to 
$z \sim$ 6--8 galaxies recently observed with the new WFC3 camera onboard HST
and other, more luminous galaxies. We have shown an overall trend of the star-formation rate
(SFR) increasing with stellar mass, albeit with a large scatter. Also, we have shown that
the specific SFR may be higher than previously thought, depending on model assumptions.
We have also  found indications for dust in some galaxies at this redshift, and a possible trend of increasing dust attenuation 
with galaxy mass. However, since only few of these galaxies are detected beyond $\ga$1.6--2 \micron,
the uncertainties of the physical parameters are fairly large.
 
To improve on this aspect, and to examine systematically the effects of nebular emission
on the physical parameters of distant galaxies, we have examined a large sample
of $z \sim$ 3--6 galaxies. Some results from this study \cite[(dB11)]{dB11}  are highlighted here.

\subsection{$z \sim$ 3--6 galaxies: photometric data and selection}
We have used the GOODS-MUSIC catalogue of \cite[Santini et al.\ (2009)]{santini09} 
providing photometry in the U, \bacs, \vacs, \iacs, \zacs, J, H, K, bands
mostly from the VLT and HST, and the 3.6, 4.5, 5.8, and 8.0 \micron\ bands
from the IRAC camera onboard {\em Spitzer}.
Using standard criteria we have selected U, B, V, and i-drop 
galaxies. To reduce the contamination rate (typically $\sim$ 10--20 \%) we
have only retained the objects whose median photometric redshifts agree 
with the targetted redshift range. This leaves us with a sample of 389, 705, 199, and 60 
galaxies with median photometric redshifts of $\zphot= 3.3$, 3.9, 4.9, and 6.0.
See \cite[dB11]{dB11}  for more details.

\begin{figure}[htb]
\vspace*{-1.0 cm}
\begin{center}
 \includegraphics[width=14cm,viewport=0 80 590 400 ,clip=true]{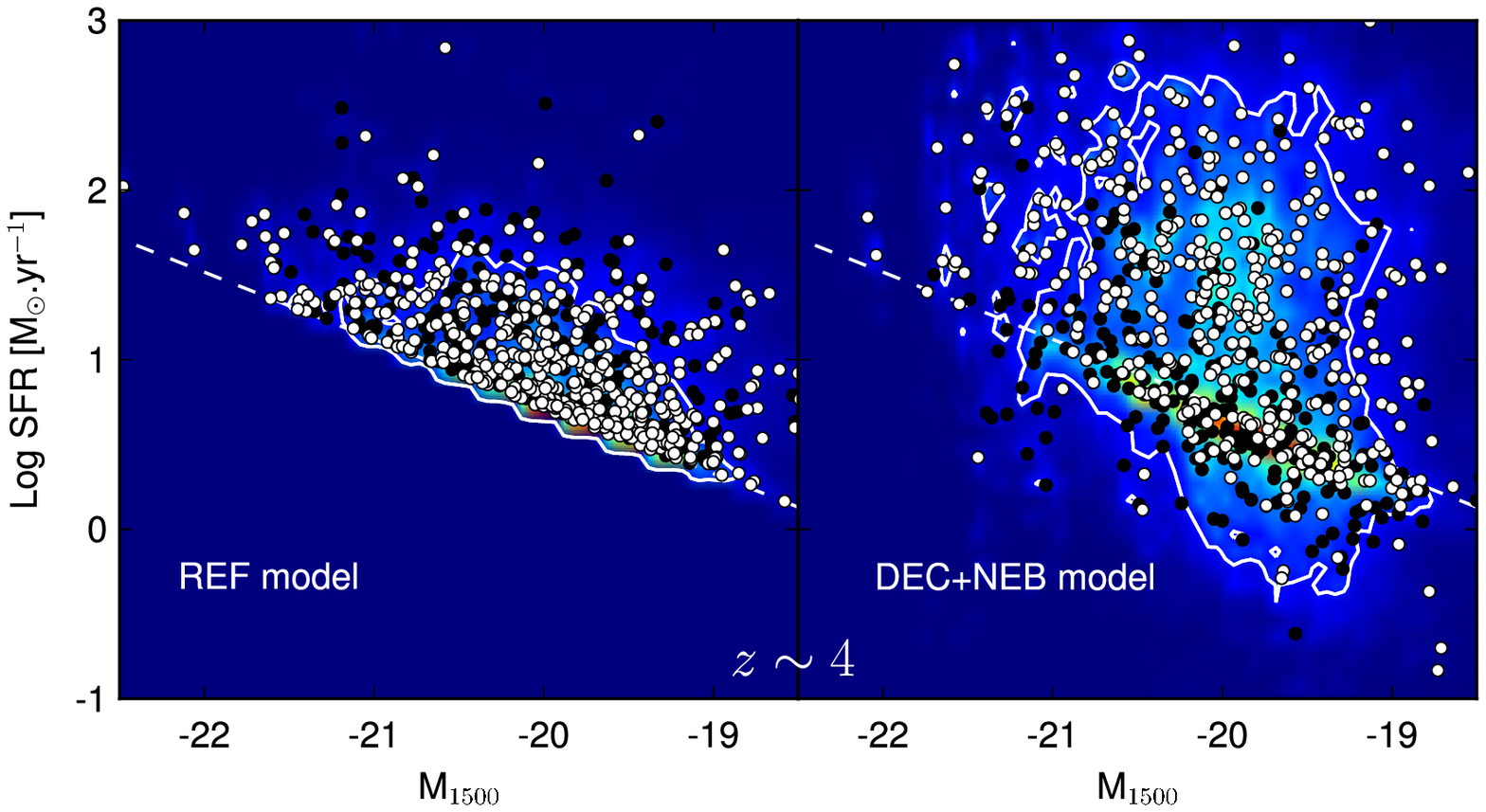} 
 \includegraphics[width=14cm,viewport=0 80 590 400 ,clip=true]{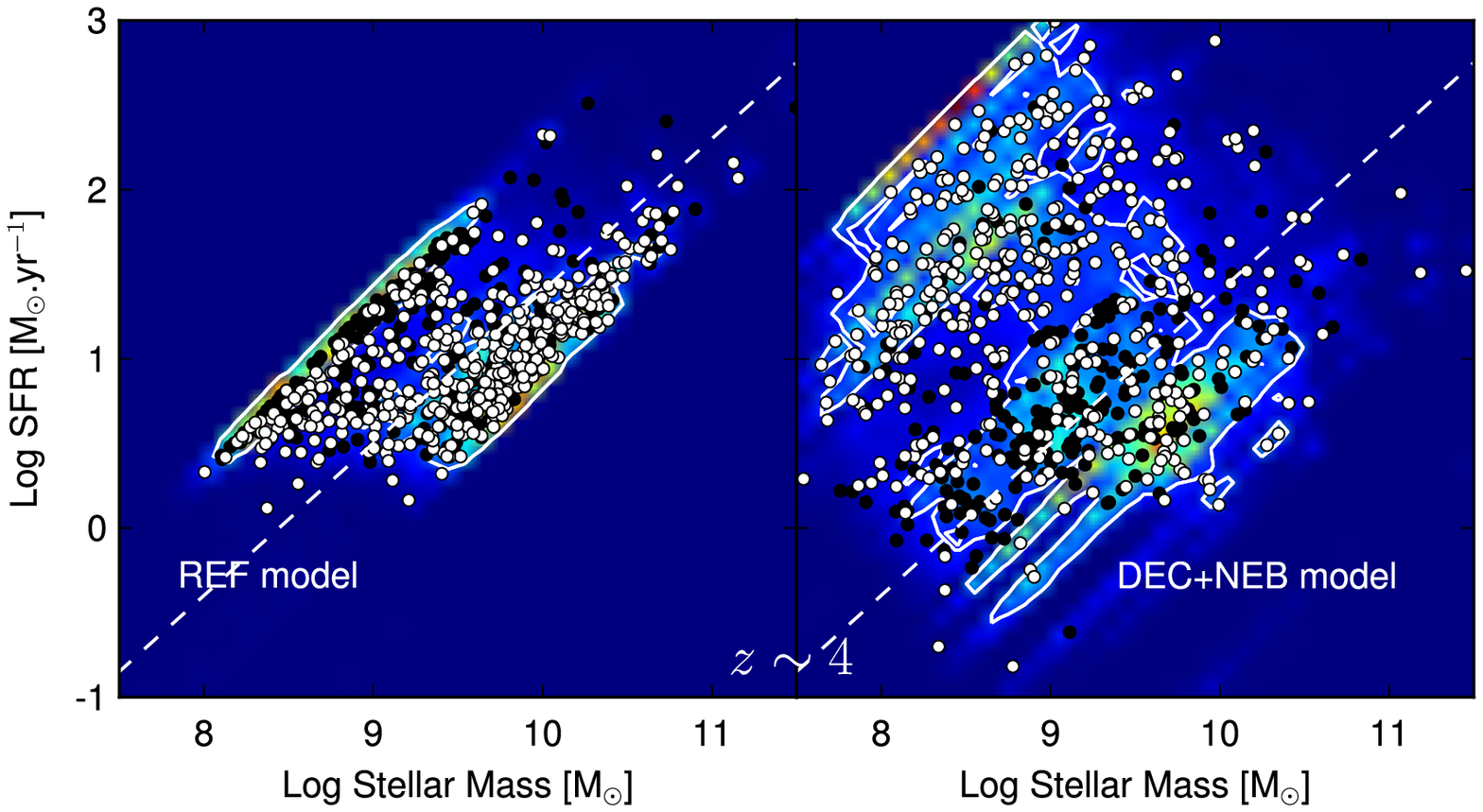} 
 \caption{SFR versus absolute UV magnitude (top) and SFR versus stellar mass (bottom) for the B-drop sample
 ($z\sim 4$) obtained from models with constant SFR without nebular emission (left panels) and
for declining SFH and nebular emission (right panels).
Points show the median values for each galaxy, color grades shows the 2D probability
distribution function, the white solid line encloses the 68 \% confidence limit.
White points show the ``starburst'' objects, black ``post-starburst'' galaxies.
The dashed lines show the SFR(UV) relation from \cite[Kennicutt (1998)]{Kenn98}
and the mean SFR-$M_\star$ relation of \cite[Daddi et al.\ (2007)]{Daddi07}.}
   \label{fig_sfr}
\end{center}
\end{figure}

\subsection{Results -- Two categories of high-z galaxies}
One of the first results obtained when applying SED models with/without nebular emission to
our large galaxy sample is that the majority of the objects ($\sim$ 60--70 \%) are better
fit with nebular emission, irrespectively of the SF history. For the rest, the difference between
fits with or without nebular emission is generally minor.
The difference between these two categories corresponds overall to objects whose
best-fit SED clearly shows nebular lines (for the majority), and weak or no lines for the rest.
Empirically this can be verified for $z\sim$ 3.8--5 galaxies, where \ha\ falls in the 3.6 \micron\
filter, whereas the 4.5 \micron\ filter remains essentially free of strong emission lines.
Indeed, for this redshift range we find that the objects best fit with nebular lines
show a systematically bluer (3.6-4.5) color than the other category \cite[(see de Barros et al. 2011b)]{dB11b}.
Our first category strongly resembles the ``\ha\ emitters" identified by \cite[Shim et al.\  (2011)]{shimetal11}
through their 3.6 \micron\ excess.
 
 From our modeling we suggest that the two categories correspond to ``starburst" and ``post-starburst"
 phases. This is supported by their difference in (instantaneous) SFR, where the objects better fit with
 emission lines (``starbursts") show on average a higher SFR than the ``post-starburst'' category 
 (see Fig.\  \ref{fig_sfr}), and also by an age difference.
 Interestingly we find a quite similar distribution of galaxies between these categories for all redshifts
 between $z \sim$ 3 and 6.

\subsection{Results -- Mass--SFR relations and the specific SFR at high-z}
The mass-SFR relation of galaxies and its evolution with redshift has drawn of lot of attention
during the last few years. Some of the issues debated include the scatter in this relation and its
behaviour at $z \gg 3$ \cite[(e.g.\ Daddi et al.\ 2007, Bouche et al.\ 2010)]{Daddi07,Bouche10}.

In Fig.\ \ref{fig_sfr}  we show the derived SFR from our $z \sim 4$ sample
as a function of UV magnitude and stellar mass respectively, and how these relations depend on
assumptions on the star-formation histories and on the treatment of nebular emission.
The reference model (left panels) assumes SFR=const and ages $\ge$ 50 Myr, as  often assumed
in other studies. The derived SFR values are found on and above the SFR(UV) calibration of 
\cite[Kennicutt (1998)]{Kenn98} indicated by the dashed line, deviations from this line being due to dust attenuation.
When nebular emission is treated and the SF history is kept free (among the declining SF histories)
both effects contribute to increasing the scatter around the Kennicutt relation (right panels).
Models with nebular emission naturally reveal a difference in SFR between the two categories 
discussed above, consistent with the observed color in the IRAC bands (3.6-4.5 \micron).

The derived stellar masses do not strongly depend on the treatment of nebular emission.
On average we obtain a reduction of $\sim$ 30--40 \% for $z \sim$ 3--4 and somewhat
more ($\sim$ 70--80 \%) at higher redshift.

As shown in Fig.\ \ref{fig_sfr} the existence of a tight mass--SFR relation relies on the assumption
of constant SFR over long enough timescales ($\gg$ 100 Myr). Again for variable SF histories and including
nebular emission we find a large scatter (e.g. around the mean relation obtained by \cite[Daddi et al.\ (2007)]{Daddi07}
at $z\sim 2$ shown in this Fig.).
The median specific SFR ($=$SFR$/M_\star$ ) of $z \sim$ 3--5 galaxies  is typically a factor $\sim$ 2--3 
higher when nebular emission is accounted for both for SFR=const and for variable SFHs
\cite[(cf.\ de Barros et al.\ 2011b)]{dB11b}.

\subsection{Results -- Dust attenuation, galaxy ages and SF histories}
\begin{figure}[htb]
\begin{center}
 \includegraphics[width=14cm,viewport=0 100 590 320 ,clip=true]{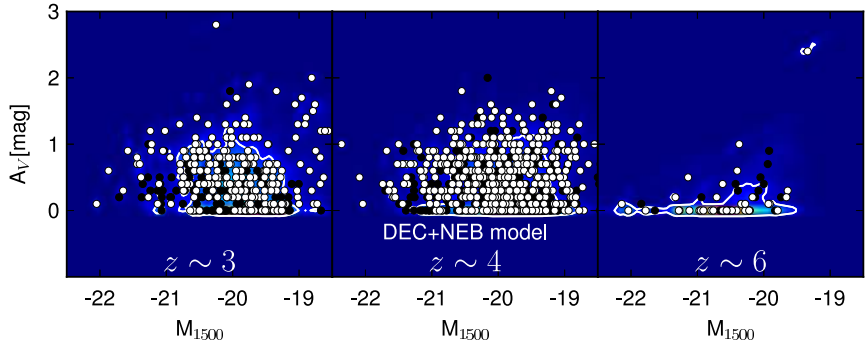} 
 \caption{Attenuation $A_V$ as a function of UV magnitude for the $z\sim$ 3 (left), 4 (middle), and 6 (right)
samples, as determined from models with declining SFHs and accounting for nebular emission.
Symbols as in Fig.\ 1. Note the fairly large scatter of $A_V$ at given $M_{\rm UV}$, and the clear decrease 
of the attenuation at high redshift.}
   \label{fig_av}
\end{center}
\end{figure}

Our analysis also yields new information on dust attenuation and on trends with UV magnitude, 
galaxy mass, redshift etc.
For illustration, we show the attenuation $A_V$ versus $M_{1500}$ for the U-, B-, and i-drop samples
in Fig.\ \ref{fig_av}. A clear evolution of decreasing $A_V$ with redshift is found, both at a given UV magnitude
and for the sample average. The exact value of the attenuation depends on model assumptions
such as the SF history and of course also on the attenuation law. Taking nebular emission into account
does not lead to a systematic shift of the inferred attenuation.

At $z \sim $ 3--4 the median age of galaxies shows a wide range from very young to close to the maximum
age allowed by cosmology. Including the effects of nebular emission yields on average somewhat younger
ages (as already found in our earlier papers), but does not preclude old ages. No strong evolution of the 
average galaxy age with redshift is found between $z \sim$ 3 and 5 e.g.\ at fixed UV magnitude or
given stellar mass; at higher $z$ our sample appear significantly younger, though. 

Interestingly declining SF histories are clearly favoured for the majority of 
galaxies, as they fit better than models assuming SFR=const or rising SF. As already mentioned above,
declining SF histories also provide a natural explanation for the two galaxy categories identified 
(``starburst" and ``post-starburst").
Other arguments, e.g.\ from galaxy statistics \cite[(Lee et al. 2009, 2011)] {Lee09,Lee11}
or others \cite[(Stark et al. 2009)]{stark09}, also tend to favour short times scales of SF.
Timescales and SF histories of high-z galaxies are, however, subject to debate \cite[(e.g.\ 
Maraston et al.\ 2010, Finlator et al.\ 2011)]{Finlator11,Maraston10}.
For more  detailed results and discussion of other implications see \cite[dB11]{dB11} .
 
\section{Conclusions}
As our previous results on $z \sim$ 6--8 galaxies \cite[(Schaerer\& de Barros 2009, 2010)]{SdB09,SdB10},
and the latest study of 
$z \sim$ 3--6 objects (\cite[dB11]{dB11}) show, it is important to include the effects of 
nebular emission in SEDs models when deriving physical parameters from broad-band
photometric data. 
SED models including nebular lines and variable \lya\ can also be used to infer
trends of \lya\ emission from photometric data, as shown in \cite[Schaerer et al.\ (2011]{Schaerer11}.

Accounting for nebular emission can lead to significant differences in the inferred
age (which tend to become younger), masses (slightly lower), SFR , 
specific star-formation rate, attenuation and other parameters.
We also find that the physical parameters depend quite significantly on the assumed
star-formation histories. Our models favour relatively short (or recurrent?) star-formation timescales.

From our analysis of a large sample of high-z galaxies we find that the majority
of objects ($\sim$ 60--70\%) is better fit with SEDs accounting for nebular emission 
(with a contribution proportionaly to the Lyman continuum flux).
This category of apparently strongly SF objects (``starburst'') can also be distinguished
empirically from the rest of the LBG population, which are probably closer to ``post-starbursts''.
We suggest that a fairly large scatter could be intrinsic in relations between SFR--mass
and others. 
More detailed results  and  implications are presented and discussed in \cite[dB11]{dB11}.


\end{document}